\documentclass[journal]{IEEEtran}

\usepackage{amssymb}
\usepackage{graphicx}
\usepackage{hyperref} 
\usepackage{amsmath}
\usepackage{tabularx}
\usepackage{amssymb}

\ifCLASSINFOpdf
\else
\fi

\begin{document}
\title{In-Device Feedback in Immersive Head-Mounted Displays for Distance Perception During Teleoperation of Unmanned Ground Vehicles}
\author{Yiming Luo$^{1}$, Jialin Wang$^{1}$, Rongkai Shi$^{1}$, Hai-Ning Liang$^{1,*}$ and Shan Luo$^{2}$
\thanks{$^{1}$Y. Luo, J. Wang, R. Shi, H.-N. Liang are with the Department of Computing, Xi'an Jiaotong-Liverpool University, Suzhou, China, 215028.}%
\thanks{$^{2}$S. Luo is with the Department of Computer Science, The University of Liverpool, Liverpool, UK, and with the Department of Engineering, King's College London, London, UK, WC2R 2LS.}%
\thanks{$^{*}$Corresponding author ({\tt\small haining.liang@xjtlu.edu.cn}).}%
}

\markboth{Journal of \LaTeX\ Class Files,~Vol.~14, No.~8, August~2015}%
{Shell \MakeLowercase{\textit{et al.}}: Bare Demo of IEEEtran.cls for IEEE Journals}

\maketitle
\begin{abstract}
 In recent years, Virtual Reality (VR) Head-Mounted Displays (HMD) have been used to provide an immersive, first-person view in real-time for the remote-control of Unmanned Ground Vehicles (UGV). One critical issue is that it is challenging to perceive the distance of obstacles surrounding the vehicle from 2D views in the HMD, which deteriorates the control of UGV. Conventional distance indicators used in HMD take up screen space which leads clutter on the display and can further reduce situation awareness of the physical environment. To address the issue, in this paper we propose off-screen in-device feedback using vibro-tactile and/or light-visual cues to provide real-time distance information for the remote control of UGV. Results from a study show a significantly better performance with either feedback type, reduced workload and improved usability in a driving task that requires continuous perception of the distance between the UGV and its environmental objects or obstacles. Our findings show a solid case for in-device vibro-tactile and/or light-visual feedback to support remote operation of UGVs that highly relies on distance perception of objects.
\end{abstract}


\IEEEpeerreviewmaketitle

\section{Introduction}
Interest in improving real-time remote human-robot interaction is growing rapidly. First-person view (FPV) using VR for unmanned ground vehicle (UGV) with remote or (semi-) automatic control is increasingly used for search and rescue operations, in disaster recovery, and for terrain and object surveillance, especially in unsafe environments \cite{a2,a4}. Immersive VR displays for UGV teleoperation can improve the user's concentration and performance in obstacles avoidance tasks when compared to a normal display such as desktop monitor \cite{b56}. UGV can explore the surrounding environment and send information captured via sensors or cameras to remotely located users in real-time. However, most cameras attached to these robots have limitations, such as low degrees of freedom, narrow field-of-view, and poor photo-sensitivity, especially in dark and complex environments with interference from the objects or obstacles found in such environments. 

\begin{figure}[t]
  \centering
  \includegraphics[width=0.8\linewidth]{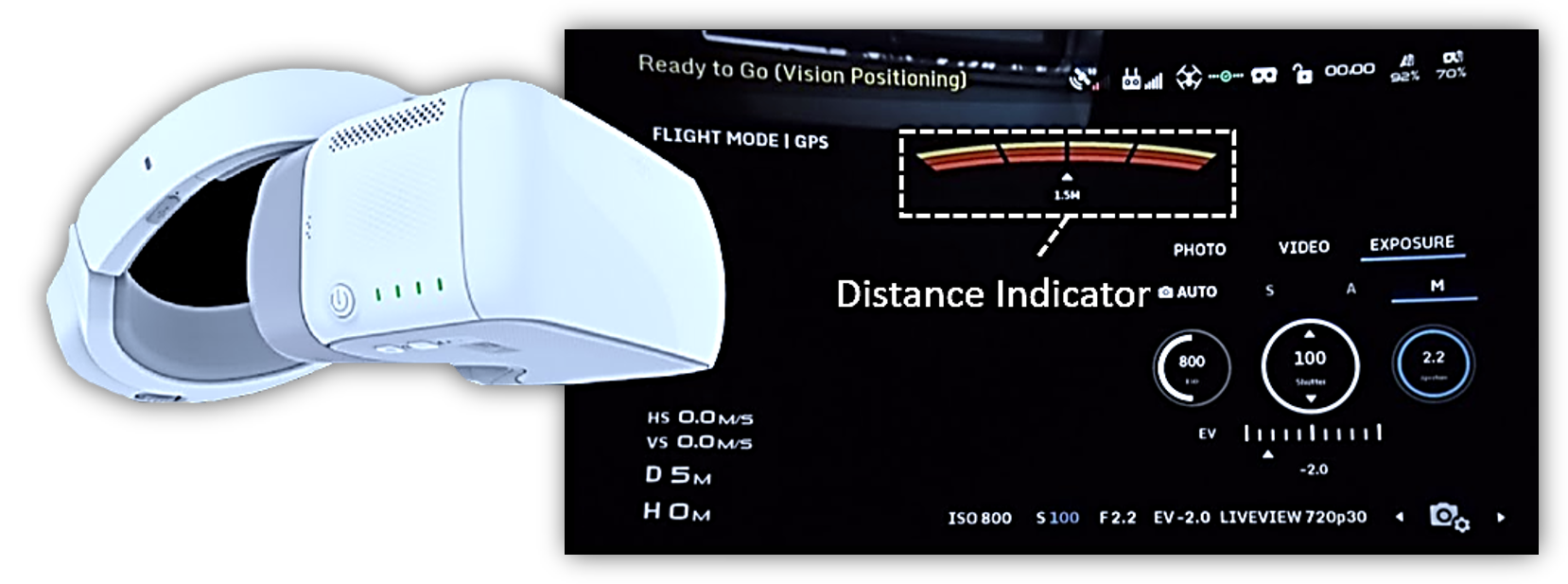}
  \caption{Interface from DJI VR HMD. Its interface for teleoperation uses visual elements on the display to provide distance information, which occupy part of the screen.}
  \label{DJI}
\end{figure}

Distance perception has been studied to improve user experience and task efficiency while using an HMD. One common way for perceiving distance information when wearing an HMD is to give visual distance indicators or warning signs on the screen (e.g., DJI Goggles\footnote{DJI Goggles: \url{https://www.dji.com/dji-goggles}}, as shown in Fig.~\ref{DJI}). However, these extra visual elements take up screen space, which is often limited in HMD. The second way is to use sound feedback, but this approach is not suitable for noisy environments \cite{a36} and is not as accessible and accurate as visual elements. These issues could be addressed with a real-time feedback system for FPV UGV using tactile and off-screen visual feedback.

Current technology for vibro-tactile devices provides low-energy, wearable, and wireless components, which offer a good level of reliability \cite{a5}. Prior research has indicated that vibro-tactile feedback could be employed in distance judgment and alerting people of incoming dangers (e.g., \cite{a6,a7,a8}). It can also improve the user's experience and sense of the environment. Users can feel the vibration feedback through direct contact with the actuators. When the user is focusing on the screen, vibro-tactile feedback can provide a second channel to assist the user in teleoperating a UGV. In addition, off-screen visual feedback from LED lights installed on the periphery of the screen can also be used as a source of distance perception information without taking up screen space in the HMD. It is possible to have different types of sensors and cameras on the UGV to allow them to scan the environment and provide more comprehensive and precise information to their remote users. This information can also be used to provide tactile and off-screen visual feedback and offer real-time and precise cues to users.

In this research, we explore the following research question: when the user operates a UGV remotely using a VR HMD, how can the user's distance perception of objects be improved without adding visual elements to the VR screen? To find answers to this, we investigated whether adding vibro-tactile and/or light-visual feedback to an HMD could improve the distance perception of users and their performance when driving a UGV remotely. We developed a VR HMD prototype to allow operating a UGV remotely. We ran a set of experiments to assess users’ driving performance, distance perception, workload demands, and preferences when driving a UGV in FPV using a VR HMD. Our results show that adding vibro-tactile feedback significantly improved operators' distance perception, reduced work demands, and increased user preferences.

The main contribution of this paper is an examination of two off-screen feedback types that can provide distance information to users when controlling a UGV remotely. Our VR HMD prototype using vibro-tactile actuators and LED lights placed inside the HMD is a simple, efficient, and low-cost approach. To the best of our knowledge, this research is the first to apply these two types of feedback inside a VR HMD to assist teleoperators in gauging the distance of objects that surrounds a UGV.

\section{Related Work}
\subsection{Distance Perception during Teleoperation}
One important issue during teleoperation, when compared to onsite operation, is the loss of perceptual and environmental information (e.g., obstacle distance) due to the limited field-of-view, low display resolution, and reduced depth perception details of objects in the environment \cite{a11,a12}. To enhance the view of the remote environment, researchers have attempted to use 3D stereoscopic views instead of 2D views by using a binocular camera to improve the depth perception and immersion level when controlling an unmanned aerial vehicle remotely (e.g., \cite{a14,a15,b41}). Their approach provides additional depth information to the operator to make it easier for the operator to estimate obstacles' distance towards the robot. Similarly, Brown et al. \cite{a16} explored the performance of a spherical camera in combination with a VR HMD to supply more intuitive multiple views and more information from the environment to improve the teleoperation process. These methods have focused on providing more visual information on the display and perception issues remain for efficient remote controllability. While they can help estimate distance information, the visual elements take up screen space and make the interface cluttered.

Instead of cameras, approaches that use of distance sensors as actuators include sonar \cite{b45,b46}, LiDAR \cite{b47}, LASER \cite{b49,b50} can improve distance perception in teleoperation. These sensors can provide more accurate distance information than through video cameras. Typically, during teleoperation, objects of interest are those close to the UGV. Sonar sensors have stable and accurate performance when detecting these near objects. In our research, sonar sensors are attached to the UGV to help detect the distance between the vehicle and its surrounding objects.

\subsection{Vibro-tactile Feedback for Immersive Displays}
Vibro-tactile feedback provided via actuators has been applied to a variety of devices, such as mobile phones or game controllers \cite{a20,a21}, to communicate information about incoming text messages and give warning notices of a dangerous situation for visually impaired people, like an obstacle on their way \cite{a22,a1}. One important feature of this type of feedback is that it is eyes-free, allowing people to focus on other activities. In addition, it can be sometimes hands-free, which allows people to do other activities with their hands.

Other research has explored how to use haptic feedback when perceived via people’s heads \cite{a1,a27,b57}. Spatial (orientation) recognition, comfort, accuracy, and tolerable vibration frequency range are important factors for vibro-tactile devices placed inside a VR HMD \cite{a6,a27,b54}. The installation position and placement of the actuator are quite important because different parts of the head have varied levels of tactile perception. For example, results in \cite{a27} show that the hairless skin on people’s forehead is the most accurate area for spatial discrimination. On the other hand, the occipital and temple regions of the head are most sensitive to vibration \cite{a29}. When it comes to using vibratory feedback on the head, the vibration frequency is also important. Research has shown that users feel discomfort when the vibration frequency is above 150 Hz while the appropriate frequency is 32 Hz \cite{a27}. The next factor that has an impact is the number of vibration motors. The actuator density influences the performance when it comes to obstacle detection or navigation tasks using vibro-tactile feedback \cite{b54}. It has been shown that there is a negative correlation between increased reaction time and decreased spatial resolution when there are higher numbers of motors \cite{a30}. In short, based on this prior research, to be effective for VR HMD, the motor(s) should be placed inside the headset and closer to users’ skin to maximize contact. 

\subsection{Light-visual Feedback for Immersive Displays}

Using light for feedback inside VR HMD has also been explored (e.g., in \cite{a31,b58}). Unlike visual feedback shown on the display directly, using LED light strips placed inside the HMD and around the two displays could also be used to give users an off-screen non-visual cluttered way to receive feedback. Xiao and Benko \cite{a31} presented \textit{SparseLight}, a prototype with a matrix of LED placed in an HMD to create higher immersive experiences. Results of their experiment show that \textit{SparseLight} is helpful in conveying peripheral information, improving situational awareness, and reducing motion sickness \cite{b59}. Gruenefeld et al. \cite{a32} imitated \textit{SparseLight} and made two ring-shaped LED matrices placed on the two circular screens of the HMD to indicate direction information and help with object selection tasks. Given their usefulness in conveying information in an eyes-free manner, we would also like to explore how effective light is to communicate distance information of objects surrounding a UGV. As such, in addition to vibro-tactile actuators, our HMD prototype also incorporates strips of LED lights. Researchers in \cite{b53} employed both light and vibro-tactile for feedback in an alarm system integrated into a multi-modal HMD. The prototype they made inspired us to design our prototype in the placement of actuators and stimulation patterns.

\section{User Study}
\begin{figure*}[t]
  \centering
  \includegraphics[width=0.9\linewidth]{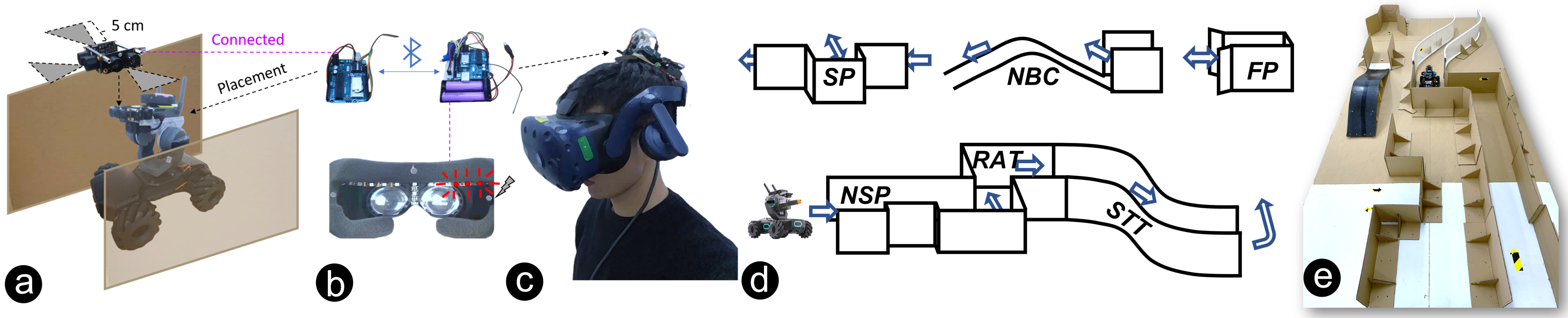}
  \caption{The components of the VR HMD prototype. (a) Three distance sensors were placed on the UGV; the detection range was set to be 5 cm; (b) upper part: the communication and processing blocks on UGV and HMD; they were connected through a Bluetooth module; lower part: the placement of actuators (coin motors and LED strips); and (c) the placement of the block on the HMD worn by a user; the block was relatively light and its placement on the HMD was designed to achieve balance with the weight of the front part of the device; (d) A diagram of the experimental site with the six local areas (tasks) -- NSP, RAT, STT, FP, NBC and SP; (e) a photograph of how the actual site looks like. The two whiteboards represent the start and end areas.}
  \label{Apparatus}
\end{figure*}
\begin{table}[t]
\caption{The metrics of the two feedback types (vibration and light)}
\label{table1}
\begin{center}
\begin{tabular}{|l|l|l|}
\hline
Distance (d)/ cm                & Vibration/ PWM level & Light/ lit LEDs \\ \hline
[0, 1, 2, 3, 4, 5] & [155, 124, 93, 62, 31, 0]    & [5, 4, 3, 2, 1, 0]          \\ \hline
\textgreater 5          & 0             & 0                   \\ \hline
\end{tabular}
\end{center}
\end{table}

In this study, we aimed to explore whether in-device feedback, as an off-screen and non-auditory secondary channel, is helpful to provide efficient, simple, and practical distance information to users when controlling a UGV. The prototype used in this study offers two feedback types based on LED lights and vibro-tactile actuators. Thus, we identified two variables (with-without light-visual feedback $\times$ with-without vibro-tactile feedback), which led to four conditions in total: (1) No feedback, Baseline (\textit{N}); (2) Light-visual only (\textit{L}); (3) Vibro-tactile only (\textit{V}); (4) Light-visual + Vibro-tactile (\textit{LV}).

This study involved 16 participants (12 males and 4 females, aged between 19--28, $M=20.5$) and followed a within-subjects approach, where all participants experienced all four conditions counterbalanced using a Latin square design to reduce the learning effects. Data from the pre-experiment questionnaire showed that it was the first time for all 16 participants to operate a remotely-controlled UGV using an HMD in FPV. Interviews with participants revealed that none of them had significant physical discomfort, health problems, simulator sickness, or vision issues. All of them were able to complete the pre-training session successfully.

\subsection{Apparatus}
A DJI RoboMaster S1\footnote{DJI RoboMaster S1: \url{https://www.dji.com/robomaster-s1}} was used as the mobile UGV. The FPV images from the camera were transmitted to a VR HMD, which in our case was an HTC Vive Pro Eye\footnote{HTC Vive Pro:\url{ https://www.vive.com/uk/product/vive-pro-eye/specs/}}. The HMD was connected to a desktop with 16GB RAM, an i7-9700k CPU, a GeForce GTX 2080Ti GPU. Participants used an Xbox wireless controller as the input device to control the UGV. 

Fig.~\ref{Apparatus}a shows the UGV prototype used in our experiment. We attached three ultrasonic range sensors that can detect distances from 2cm to 500cm with low latency. We placed the sensors on the car where they were stable while the car was moving and were able to sense its surroundings with precision.

There were two communication and processing blocks (see the upper part of Fig.~\ref{Apparatus}b). Each of them consists of one Arduino Uno micro-controller and one Bluetooth module. The two blocks were placed on the UGV and the HMD, respectively. The camera's images captured from the UGV were transmitted in real-time to the HMD via an image transmission unit.

According to \cite{a1}, when the voltage on the vibration motor is changed, we can manipulate the speed of its rotor to control its vibration intensity. It is important to note that in coin motors the vibration amplitude and vibration frequency are closely linked. Each motor is controlled by pulse-width modulation (PWM) signal from the processing block. Therefore, the haptic stimulation changes with driving voltage. In order to make the haptic stimulation within the acceptable range, we limited the diving voltage to a certain range ([0, 3V] with PWM range: [0, 155]) to meet the requirements of the vibration frequency mentioned in \cite{a27}. As for visual stimulation, we chose the number of lit LEDs to present the light intensity. To compare the two feedback types, we show the metrics in Table~\ref{table1}. The lower part of Fig.~\ref{Apparatus}b shows the final design, with the placement of the vibration motors and LED strips inside the HMD. Their placement was determined by a pre-experiment with 8 pilot users. Our main requirement was for the items to be in the hairless area of the front and occipital regions \cite{a16}. The pre-experiment sessions also helped determine the placement of the LED strips, which was inspired and adapted from \cite{a31}. In addition, the maximum comfort intensity for the vibration and LED light strength was also determined in this pre-experiment. Our design of these two off-screen feedback mechanisms was efficient and practical, and allowed for ease of integration into the HMD without increasing discomfort.

In our design, the intensity of stimulation increases as the UGV approaches the obstacles. According to \cite{a6}, the transfer functions (Step, Linear, Gamma, Linear-step, Gamma-step) for mapping the distance between obstacles and sensor to a vibration level of an actuator did not show significant effects on the number of collisions and completion time. As such, we did not evaluate the effect of different transfer functions. Since the main purpose of the vibration feedback is to provide distance information, the stimulation should be strong at a very close distance, which results in a linearly decreasing function.

\subsection{Tasks and Procedure}
To investigate the performance of the two different types of feedback, we asked participants to drive the robot through a customized test site (see Fig.~\ref{Apparatus}e). The site had walls made of cardboard along the navigation path because we were interested in observing the effect of the sensors in providing dynamic information to users as the distance between the car and the walls changed in real-time.

Participants had to wear the HMD prototype and drive the UGV remotely. The driving scenario was inspired by elements of actual driving tests. The main purpose of this test is to complete a series of tasks such as reversing or turning, while the UGV is not allowed to be outside of the designated areas. The requirement of our experiment is to avoid collisions with the walls and complete the prescribed tasks in the shortest time possible. The number of collisions and completion time was video-recorded with two high-definition video cameras for later analysis.

A simple pre-training driving session was given to the participants before they started the formal trials. The purpose of this training was to get participants familiar with the controllers, the HMD and each feedback mechanism, and controlling the robot. After this training, the participants were asked to run the formal trials. Each participant had four trials, one for each condition, whose order was pre-determined and counterbalanced to minimize any learning effects.

There were six tasks involved in the driving test (see Fig.~\ref{Apparatus}d) that were mapped to the site. (1) Narrow straight passage (\textit{NSP}): The UGV needed to pass through a narrow and straight passage with different width; (2) Right-angle turn (\textit{RAT}): The UGV had to make a right-angle turn; (3) S-type turn (\textit{STT}): The UGV needed to make an S-type turn; (4) Forward parking (\textit{FP}): The UGV had to go inside an enclosure and make a full stop to park in a forward direction; (5) Narrow bridge crossing (\textit{NBC}): The UGV had to cross a narrow bridge; and (6) Side parking (\textit{SP}): The UGV needed to be parked sideways in the enclosure. After each condition, the participants were given a break and completed the NASA-TLX workload questionnaire \cite{nasatlx} and User Experience Questionnaire (UEQ) \cite{UEQ} for the condition they just experienced. 

\subsection{Hypotheses}
Based on our review of the literature and experiment design, we formulated the following three two-part hypotheses.
\begin{itemize}
    \item \textit{H$_1$}: with light-visual feedback would lead to less number of collisions than no feedback; with vibro-tactile feedback would lead to less number of collisions than no feedback;
    \item \textit{H$_2$}: with light-visual feedback would lead to longer completion time than no feedback; with vibro-tactile feedback would lead to longer completion time than no feedback;
    \item \textit{H$_3$}: with light-visual feedback would lead to lower simulator sickness than no feedback; with vibro-tactile feedback would lead to less workload demands than no feedback.
\end{itemize}

\section{Results \& Discussion}
All participants understood the nature of the tasks, and all recorded data was valid. We discuss the results in terms of participants' performance in six tasks in each condition. A Shapiro-Wilk test was first performed on objective measures separately for each condition. All the objective data followed normal distributions ($p>.05$). Thus, we conducted two-way repeated measure ANOVAs (RM-ANOVAs) for the comparison of objective measurements. We applied Aligned Rank Transform (ART) \cite{b55} on NASA-TLX and UEQ data before performing RM-ANOVAs on them. To improve the readability, we mainly report significant results in this section. Post-hoc tests were run with Bonferroni corrections if a significant difference was found. 

\begin{figure}[t]
  \centering
  \includegraphics[width=0.9\linewidth]{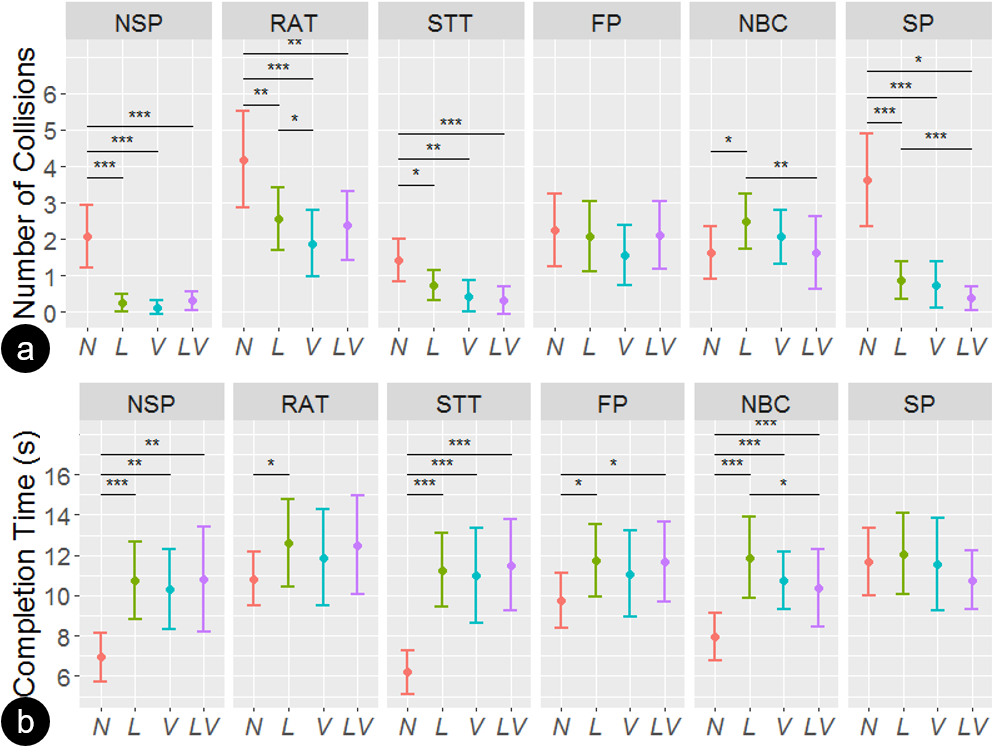}
  \caption{(a) Mean number of collisions, and (b) mean completion time (s) of each task. The error bars represent 95\% confidence intervals. `***', `**', and `*' represent a `.001', `.01', and `.05' significance levels, respectively. \textit{N}: No feedback, Baseline; \textit{L}: Light-visual only; \textit{V}: Vibro-tactile only; and \textit{LV}: Light-visual + Vibro-tactile.}
  \label{Results}
\end{figure}

\subsection{Objective Results}
The number of collisions for each trial was collected by an experimenter frame by frame from the video and validated by another experimenter. Based on our observations, the collisions in our experiment could be either discrete or continuous. A discrete collision happened when the participant immediately controlled the UGV to leave the wall, which commonly took around 0.5 seconds. We considered each discrete collision as one collision. On the other hand, while less frequent, if the participant did not adjust UGV's direction immediately after colliding with any part of a wall and continuously letting the UGV hit the wall, we considered this series of collisions as continuous collision, in which every 0.5 seconds of it was counted as one collision. 

RM-ANOVA tests revealed the mean number of collisions differed significantly for with-without light-visual feedback in \textit{NSP} ($p<.05$), \textit{RAT} ($p<.05$), \textit{STT} ($p<.001$) and \textit{SP} ($p<.001$); and for with-without vibro-tactile feedback in \textit{NSP} ($p<.001$), \textit{RAT} ($p<.001$), \textit{STT} ($p<.001$) and \textit{SP} ($p<.001$). In addition, there was also a significant interaction effect in \textit{NSP} ($p<.001$), \textit{RAT} ($p<.001$), \textit{NBC} ($p<.01$) and \textit{SP} ($p<.001$). 

Fig.~\ref{Results}a summarizes the mean number of collisions for each task in each condition and the significant differences derived from post-hoc tests. It showed that having light-visual feedback only, having vibro-tactile feedback only, and having both significantly reduced the number of collisions compared to no feedback (baseline) in \textit{NSP}, \textit{RAT}, \textit{STT} and \textit{SP}. The results supported {\textit{H$_{1}$}}. These tasks require participants to estimate the distance for both sides of UGV to obstacles. If there were only images from the screen, it would be difficult for users to judge distance because the information would be out of their field-of-view. With our prototype, the extra feedback could supply additional sensorial details for users and improve their distance perception. However, the number of collisions did not significantly change between having light-visual and vibro-tactile feedback together and having one of them only. The interaction effect only occurred when one of the two feedback types was provided. As such, we can only confirm that the two types of feedback could improve the user's distance perception by supplying additional information to users efficiently and dynamically when they work individually. Both of the two feedback mechanisms can provide precise and rapid distance information.

Results of RM-ANOVA test showed that the mean completion time differed significantly for with-without light-visual feedback in \textit{NSP} ($p<.01$), \textit{RAT} ($p<.05$), \textit{STT} ($p<.001$), \textit{NBC} ($p<.05$) and \textit{FP} ($p<.001$); and for with-without vibro-tactile feedback in \textit{NSP} ($p<.05$) and \textit{SP} ($p<.05$); In addition, we also found an interaction effect in \textit{NSP} ($p<.01$), \textit{STT} ($p<.001$) and \textit{NBC} ($p<.001$). The results from post-hoc tests were summarized in Fig.~\ref{Results}b. 

As shown in Fig.~\ref{Results}b, having light-visual feedback only, having vibro-tactile feedback only, and having both of them significantly increase the completion time in three tasks (\textit{NSP}, \textit{STT} and, \textit{NBC}) compared to no feedback (baseline) in different ways -- {\textit{H$_{2}$}} is largely supported. We had expected participants to take a longer time for two reasons. One is to be accustomed and be comfortable with the two types of feedback and internalize their use, and the second is the additional time to apply the changes to the UGV when receiving the feedback information in real-time. It is important to note that our participants had limited training prior to the experiment and were still able to use the feedback types effectively. Based on data collected from the interview, they said that with more practice, they would be more efficient and faster.

\subsection{Subjective Results}
RM-ANOVA tests with ART data revealed that having light-visual feedback ($p<.05$) significantly reduced five elements of NASA-TLX workload demands (Mental, $p<.001$; Physical, $p<.001$; Temporal, $p<.001$; Performance, $p<.001$; Frustration, $p<.05$). Having vibro-tactile feedback reduced three demands (Performance, $p<.001$; Effort, $p<.01$; and Frustration, $p<.05$). {\textit{H$_{3}$}} is supported to a large extent. It can be seen from the above results that light-visual feedback can reduce more aspects of workload demands than vibro-tactile feedback does. Although vibro-tactile feedback can improve users' confidence in performance, it did not lead to a significant reduction in mental and physical stress.

Another RM-ANOVA test to transformed UEQ results showed that having light-visual feedback only (Perspicuity, $p<.05$; and Dependability, $p<.01$) or having vibro-tactile feedback only (Dependability, $p<.001$; Stimulation, $p<.001$; and Novelty, $p<.01$) was significantly more popular with users. However, we did not find interaction effects.

According to the above results, we found that having vibro-tactile only is more creative, exciting and motivating to users compared to having light-visual feedback only. On the other hand, while light-visual feedback is less novel, it is easier for users to become familiar and comfortable with it.

\section{Limitations and Future Work}
As some participants commented, both vibro-tactile and light-visual feedback presented some usability issues, though they were minor. There is a need to address the problems with vibration producing small amounts of sound. Also, we need to eliminate the problem of the LED lights making small parts of the screen brighter, if possible.

Other types of feedback, such as non-contact haptics (ultrasound or wind) and smell, can be explored. In terms of distance perception, the smell does not provide direct and practical help. However, non-contact haptics can offer distance perception similar to vibro-tactile information while avoiding direct contact. Direct contact can cause vibration to be transmitted into the skull, which could further increase discomfort if the vibration actuator is attached too tight to the head. We plan to investigate the suitability and effectiveness of non-contact haptics on distance perception or other tasks in the future.

We confirmed that our feedback for a head is feasible to provide simple and efficient distance perception when remotely controlling the UGV, but we still need to consider some issues. In a complex environment that a UGV is navigating in general, the requirements of the accuracy of the distance information are higher, which increases the difficulty of our simple distance perception feedback to provide enough information with limited actuators. Due to the resolution of the perception of human head is limited, the number of the actuators cannot be large \cite{a30}. Therefore, we recommend using our technology under the premise of fewer actuators and rapid response. When more information is needed, it is still recommended to obtain the information on the screen. 

When there are primary and secondary tasks at the same time, the cognitive load could increase. In the future, we also plan to investigate this by evaluating the cognitive load level for each feedback and the combination of them.

\section{Conclusion}
This paper presented an in-device feedback system for distance perception for unmanned ground vehicles (UGV) controlled remotely via a virtual reality head-mounted display (VR HMD). It uses on-vehicle sensors and vibro-actuators and LED lights placed inside the HMD to provide distance information of obstacles around the UGV. Results from a driving task experiment show that the additional feedback allowed for effective navigation, with users being able to detect the distance with relatively high precision between the UGV and surrounding obstacles in real-time. This additional sensory information allowed users to move the UGV in a more precise manner and led to a more positive experience. While both vibro-tactile and light-visual feedback are helpful, each of them has its advantages. Light-visual is easier to get accustomed to using, while vibro-tactile feedback seems to provide higher accuracy, is more exciting and novel to use, and is more resistant to interference under complex driving conditions. In short, as our results show, in-device feedback using vibro-tactile actuators and LED lights is a simple, efficient, and low-cost approach to provide precise distance information for UGVs controlled via a VR HMD in real-time.

\section*{Acknowledgment}
The authors would like to thank the participants who joined the study and the reviewers for their insightful comments that helped to improve our paper. This work was supported in part by Xi'an Jiaotong-Liverpool
University--Key Special Fund (\#KSF-A-03), the Future Network Scientific Research Fund (\#FNSRFP-2021-YB-41), and Engineering and Physical Sciences Research Council (\#EP/T033517/1).

\ifCLASSOPTIONcaptionsoff
  \newpage
\fi

\bibliographystyle{IEEEtran}
\bibliography{Reference}


\begin{IEEEbiography}
[{\includegraphics[width=1in,height=1.25in,clip,keepaspectratio]{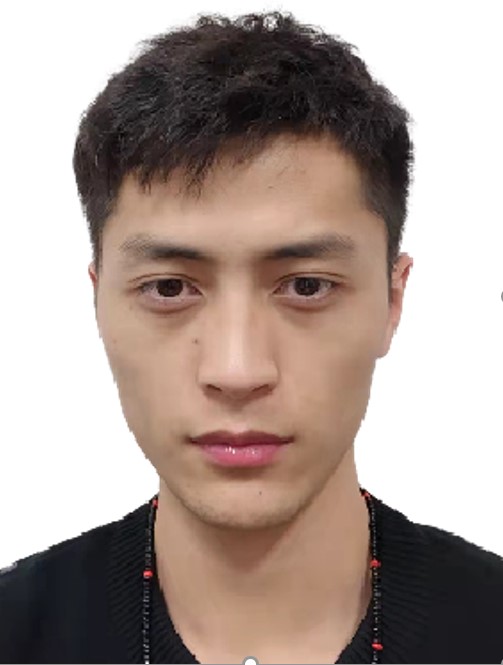}}]{Yiming Luo} is currently a PhD student in the Department of Computing at Xi’an Jiaotong-Liverpool University, China. His research interests focus on robotics, human-robot interaction, and virtual reality.
\end{IEEEbiography}

\begin{IEEEbiography}
[{\includegraphics[width=1in,height=1.25in,clip,keepaspectratio]{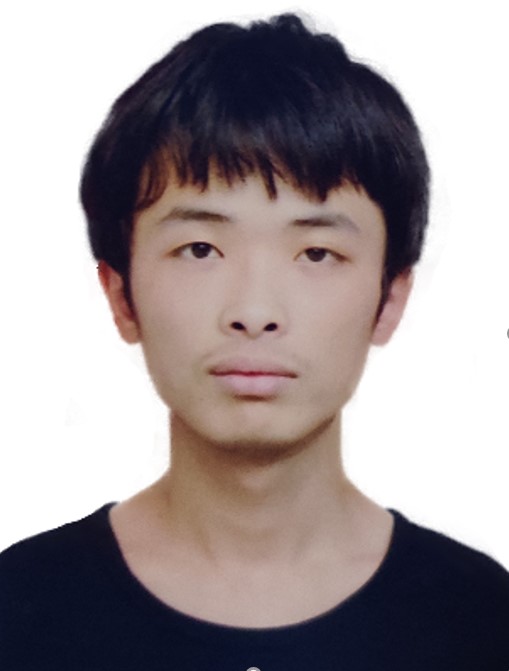}}]{Jialin Wang} is currently a first year PhD student in the Department of Computing at Xi’an Jiaotong-Liverpool University, China. His research interests include virtual reality, robotics, and computer graphics.
\end{IEEEbiography}

\begin{IEEEbiography}
[{\includegraphics[width=1in,height=1.25in,clip,keepaspectratio]{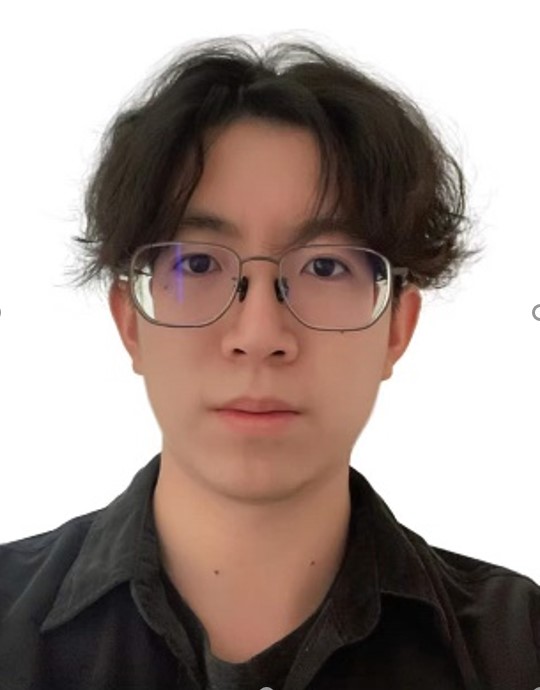}}]{Rongkai Shi} is currently a PhD student in the Department of Computing at Xi’an Jiaotong-Liverpool University, China. His research interests focus on virtual reality, augmented reality, and interaction design.
\end{IEEEbiography}

\begin{IEEEbiography}
[{\includegraphics[width=1in,height=1.25in,clip,keepaspectratio]{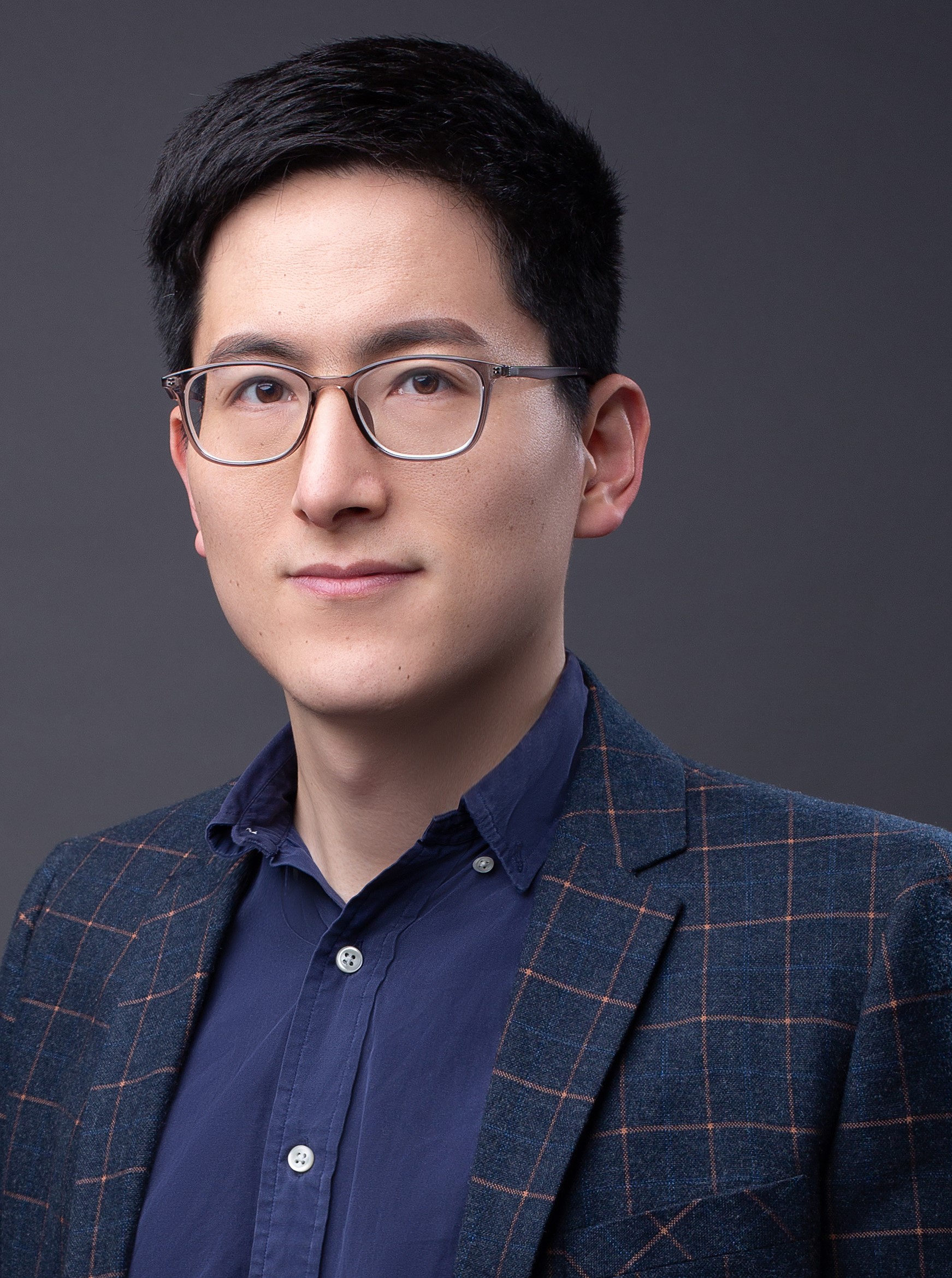}}]{Shan Luo} is Senior Lecturer (Associate Professor) at the Department of Engineering, King’s College London. Previously, he was Lecturer at the University of Liverpool, and Research Fellow at Harvard University and University of Leeds. He was also a Visiting Scientist at the Computer Science and Artificial Intelligence Laboratory (CSAIL), MIT. He received the BEng degree in Automatic Control from China University of Petroleum, Qingdao, China, in 2012. He was awarded the PhD degree in Robotics from King’s College London, UK, in 2016. His research interests include tactile sensing, robot learning and robot visual-tactile perception.
\end{IEEEbiography}

\begin{IEEEbiography}
[{\includegraphics[width=1in,height=1.25in,clip,keepaspectratio]{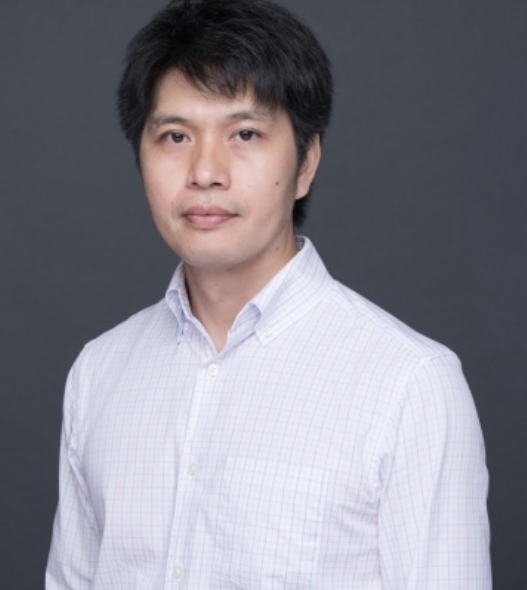}}]{Hai-Ning Liang} is Professor of Computing at Xi'an Jiaotong-Liverpool University, China, where is also Head of the Department of Computing. He received his PhD in computer science from Western University, Canada. His main research interests fall in the area of human-computer interaction, focusing on virtual/augmented reality, visualization, and gaming technologies. He has published widely in highly rated journals and conferences in these areas, such as ACM ToG, ACM ToCHI, ACM IMWUT, IEEE TVCG, ACM CHI, IEEE VR, IEEE ISMAR.
\end{IEEEbiography}

\end{document}